\documentclass{article}

\usepackage{arxiv}
\usepackage[utf8]{inputenc} 
\usepackage[T1]{fontenc}    
\usepackage{hyperref}       
\usepackage{url}            
\usepackage{booktabs}       
\usepackage{amsfonts}       
\usepackage{nicefrac}       
\usepackage{microtype}      
\usepackage{lipsum}		
\usepackage{graphicx}
\usepackage{doi}
\usepackage[nolist]{acronym}
\usepackage{amsmath}
\usepackage{tikz}
\usepackage{pgfplots}
\pgfplotsset{compat=1.18}
\usepackage{pgfgantt}

\title{Measurements of the Safety Function Response Time on a Private 5G and IO-Link Wireless Testbed}

\author{ \href{https://orcid.org/0000-0002-5390-3946}{\includegraphics[scale=0.06]{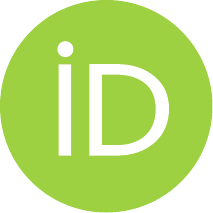}\hspace{1mm}Henry Beuster}\\
	Electrical Measurement Engineering\\
	Helmut-Schmidt-University\\
	Hamburg, Germany\\
	\texttt{henry.beuster@hsu-hh.de} \\
	\And
	\href{https://orcid.org/0009-0006-0744-9654}{\includegraphics[scale=0.06]{orcid.pdf}\hspace{1mm}Kevin Tebbe}\\
	Electrical Measurement Engineering\\
	Helmut-Schmidt-University\\
	Hamburg, Germany\\
	\texttt{kevin.tebbe@hsu-hh.de} \\
    \And
	\href{https://orcid.org/0000-0002-6882-1214}{\includegraphics[scale=0.06]{orcid.pdf}\hspace{1mm}Thomas Doebbert}\\
	Electrical Measurement Engineering\\
	Helmut-Schmidt-University\\
	Hamburg, Germany\\
	\texttt{thomas.doebbert@hsu-hh.de} \\
    \And
	{\hspace{1mm}Gerd Scholl}\\
	Electrical Measurement Engineering\\
	Helmut-Schmidt-University\\
	Hamburg, Germany\\
	\texttt{gerd.scholl@hsu-hh.de} \\
}

\date{}

\renewcommand{\shorttitle}{SFRT on a 5G and IOLW Testbed}

\hypersetup{
pdftitle={Measurements of the Safety Function Response Time on a Private 5G and IO-Link Wireless Testbed},
pdfsubject={5G, IO-Link Wireless, safety function response time, wireless framework},
pdfauthor={Henry Beuster},
pdfkeywords={5G, IO-Link Wireless, safety function response time, wireless framework},
}

\begin{document}
\maketitle

\begin{acronym}
	\acro{iolw}[IOLW]{IO-Link Wireless}
	\acro{plc}[PLC]{programmable logic controller}
	\acro{sfrt}[SFRT]{safety function response time}
	\acro{ism}[ISM]{industrial, scientific and medical}
	\acro{mmtc}[mMTC]{massive machine type communications}
	\acro{iol}[IOL]{IO-Link}
	\acro{lte}[LTE]{long-term evolution}
	\acro{ue}[UE]{user equipment}
	\acro{nsa}[NSA]{non standalone}
	\acro{sa}[SA]{standalone}
	\acro{scs}[SCS]{subcarrier spacing}
	\acro{ofdm}[OFDM]{orthogonal frequency-division multiplexing}
	\acro{urllc}[URLLC]{ultra reliable low latency communications}
	\acro{embb}[eMBB]{enhanced mobile broadband}
	\acro{revpi}[RevPi]{Revolution Pi}
	\acro{vpn}[VPN]{virtual private network}
	\acro{dhcp}[DHCP]{Dynamic Host Configuration Protocol}
	\acro{ip}[IP]{Internet Protocol}
	\acro{rssi}[RSSI]{Received Signal Strength Indicator}
	\acro{iols}[IOLS]{IO-Link Safety}
\end{acronym}

\begin{abstract}
\footnote{© 2024 IEEE. Personal use of this material is permitted. Permission from IEEE must be obtained for all other uses, in any current or future media, including reprinting/republishing this material for advertising or promotional purposes, creating new collective works, for resale or
redistribution to servers or lists, or reuse of any copyrighted component of this work in other works.} In the past few years, there has been a growing significance of interactions between human workers and automated systems throughout the factory floor. Wherever static or mobile robots, such as automated guided vehicles, operate autonomously, a protected environment for personnel and machines must be provided by, e.g., safe, deterministic and low-latency technologies. Another trend in this area is the increased use of wireless communication, offering a high flexibility, modularity, and reduced installation and maintenance efforts. This work presents a testbed implementation that integrates a wireless framework, employing \ac{iolw} and a private 5G cellular network, to orchestrate a complete example process from sensors and actuators up into the edge, represented by a \ac{plc}. Latency assessments identify the system's cycle time as well as opportunities for improvement. A worst-case estimation shows the attainable \acl{sfrt} for practical applications in the context of functional safety.
\end{abstract}

\acresetall


\section{Introduction}\label{sec:intro} 
The integration of \ac{iolw} with a private 5G campus network to establish a comprehensive wireless digital sensor-to-edge cloud infrastructure within an industrial setting has been previously proposed \cite{doebbert_study_2021, cammin_concept_2023}. Combining an economically favorable technology like \ac{iolw} with the capabilities of 5G unlocks several possibilities, such as widespread wireless sensor/actuator networks and wireless I/O aggregation. Especially, the deployment of protocols within the unregulated \ac{ism} radio bands may lead to interference among different technologies, including WiFi and Bluetooth. 5G, on the other hand, with a configuration designed for \ac{mmtc}, is able to connect up to 10\textsuperscript{6} devices per km\textsuperscript{2}, e.g., in internet of things applications \cite{noauthor_minimum_2017}.

The paper is organized as follows: Section~\ref{sec:technologies} introduces the radio transmission links and protocols utilized, followed by a description of their application in the testbed setup in Section~\ref{sec:testbedarch}. The latency measurements are presented in Section~\ref{sec:measurement}. The paper concludes with Section~\ref{sec:conclusion}, where the findings are summarized and further research directions are proposed.
 
\section{Essential technologies}\label{sec:technologies}
In the following, the primary technologies, \ac{iolw} and 5G, employed in the testbed are introduced.
 
\subsection{IO-Link Wireless}\label{sec:iolw} 
\ac{iolw} \cite{noauthor_iec_2023} is an open-vendor communication standard for factory automation, developed as an extension of the proven \ac{iol} standard IEC 61131-9 \cite{noauthor_iec_2022}. It is mainly intended for sensor/actuator communication below the fieldbus level and offers bidirectional wireless communication for cyclic process data and acyclic on-request data between a W-Master and W-Device in a star-shaped topology \cite{noauthor_iec_2023, heynicke_io-link_2018}. The physical layer is based on Bluetooth Low Energy 4.2, operating on the 2.4 GHz \ac{ism} band with Gaussian frequency-shift keying modulation \cite{noauthor_iec_2023, heynicke_io-link_2018}. Block listing of individual frequency channels is implemented to improve coexistence behavior \cite{heynicke_io-link_2018, krush_coexistence_2021}. In the same cell, up to three W-Masters can operate in parallel, with each W-Master providing one to five tracks, supporting up to eight slots each and therefore up to 120 W-Devices \cite{noauthor_iec_2023, heynicke_io-link_2018}. Additionally, roaming capabilities between different W-Masters are possible \cite{rentschler_roaming_2017, beuster_design_2024}. Deterministic media access is a crucial feature, with bidirectional communication segmented into cycles and sub-cycles. A minimum cycle lasts 5\,ms, containing three sub-cycles of 1.664\,ms each. Frequency hopping takes place between the sub-cycles, with the hopping distance being greater than the typical coherence bandwidth to increase transmission robustness, which is an essential aspect for safety applications \cite{doebbert_contribution_2024}. Furthermore, this results, in conjunction with other measures, to an error probability of 10\textsuperscript{-9} \cite{noauthor_iec_2023}.

\subsection{Private 5G Campus Network}\label{sec:5g} 
5G New Radio, the successor of 4G \ac{lte}, was defined and specified as 3GPP Release 15 in 2019 \cite{noauthor_3gpp_2019}. Like its predecessor, 5G consists of the same parts, with \ac{ue} connected to a radio access network, which in turn is connected to a core network. Two architectures are specified: a \ac{nsa} architecture with a 5G access network being connected to a 4G core network, and a \ac{sa} architecture being independent of 4G \cite{noauthor_3gpp_2019}. Another major difference is the adaptable \ac{scs} in 5G on the physical layer. \ac{lte} has a fixed \ac{scs} of 15\,kHz, resulting in a bandwidth per \ac{ofdm}-symbol of 180\,kHz. In 5G, the \ac{scs} is variable between 30~-~240\,kHz, depending on the frequency range and thus gives a bandwidth per \ac{ofdm}-symbol from 360 to 2880\,kHz. The larger the \ac{scs} is set, the lower the time slot per \ac{ofdm}-symbol is. 3GPP Release 15 introduces network slicing, dividing a physical network into multiple logical networks with different purposes. The main characteristics are \ac{urllc}, \ac{embb} and \ac{mmtc}, each with specific performance goals as depicted in \cite{noauthor_minimum_2017}, e.g., \ac{urllc} aims for a reliability of 99.999\,\% with a latency of 1\,ms.

According to German regulations, private 5G frequencies are allocated within the 3.7 to 3.8\,GHz (sub-6-GHz) band, with a bandwidth of up to 100\,MHz or in the 26.5 and 27.5\,GHz band, called mmWave, with a bandwidth of up to 1\,GHz \cite{bundesnetzagentur_administrative_2023-1,bundesnetzagentur_administrative_2023}.

\section{Testbed architecture}\label{sec:testbedarch} 
\begin{figure}[tb]
	\centering
	\includegraphics[width=1\linewidth]{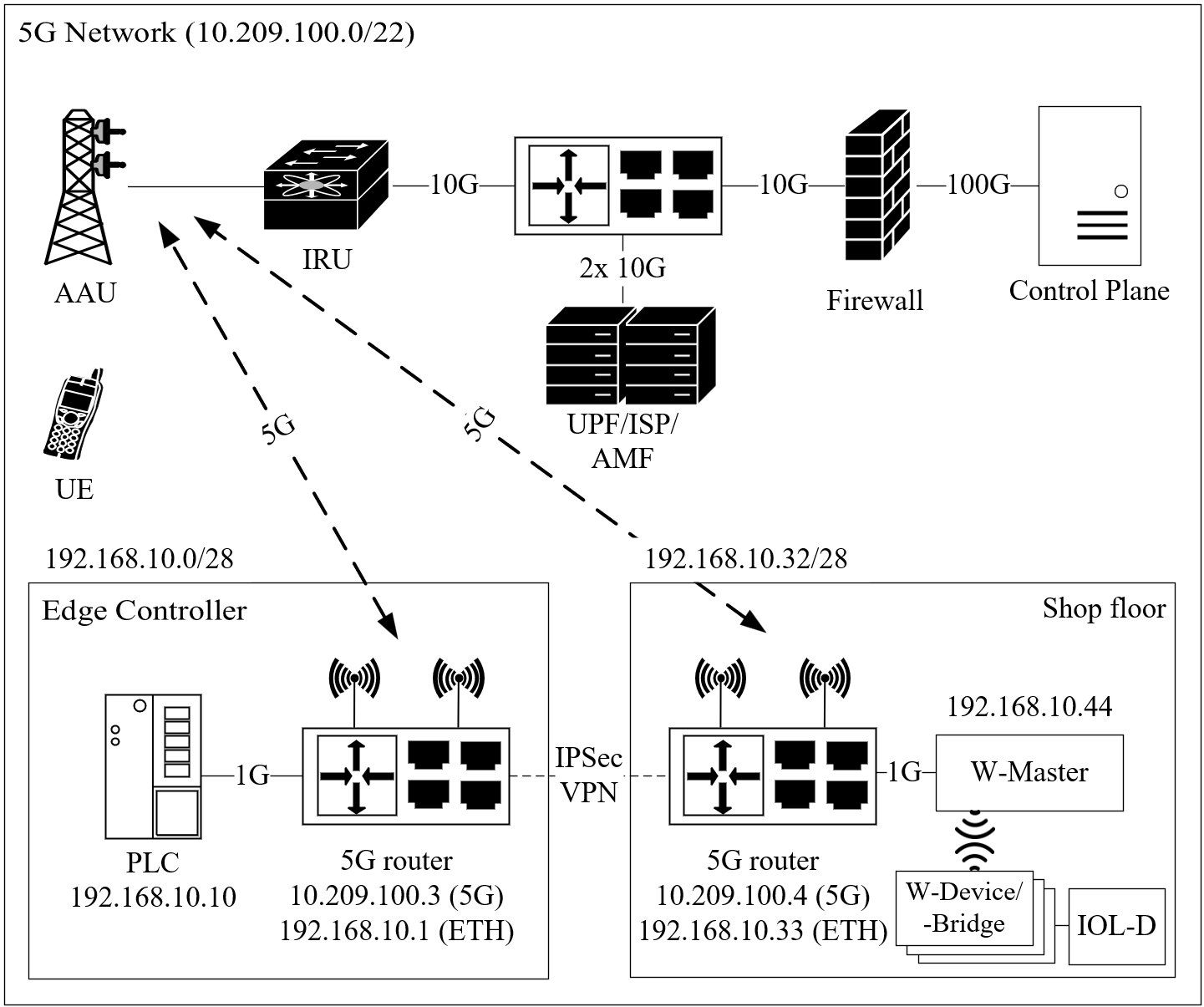}
	\caption{Testbed architecture.}
	\label{fig:testbednetworkarchitecture}
\end{figure}

The testbed emulates a production cell from an intralogistics application, segmented into two distinct but interdependent parts. An overview of the architecture is given in Fig.~\ref{fig:testbednetworkarchitecture} and Table~\ref{tab:bom} lists the main components used, the installed software and their versions.

\begin{table}[tb]
	\centering
	\caption{Bill of Material.}
	\begin{tabular}{|l|l|l|}
		\hline
		\begin{tabular}[c]{@{}l@{}}Device / \\ Function\end{tabular} & \begin{tabular}[c]{@{}l@{}}Manufacturer / \\ Software\end{tabular}                      & \begin{tabular}[c]{@{}l@{}}Model / \\ Version\end{tabular}                          \\ \hline
		5G SA network                                                & Ericsson                                                                                & EDAV-I                                                                              \\
		5G Router                                                    & \begin{tabular}[c]{@{}l@{}}Cradlepoint /\\ NetCloud OS\end{tabular}                     & \begin{tabular}[c]{@{}l@{}}R1900-5GB /\\ 7.23.110\end{tabular}                      \\
		PLC                                                          & \begin{tabular}[c]{@{}l@{}}Kunbus /\\ RevPi Buster Lite,\\ CODESYS Control\end{tabular} & \begin{tabular}[c]{@{}l@{}}RevolutionPi Connect /\\ 08/2022,\\ 4.6.0.0\end{tabular} \\
		W-Master                                                     & \begin{tabular}[c]{@{}l@{}}Kunbus,\\ IOLW Stack\end{tabular}                            & \begin{tabular}[c]{@{}l@{}}AM437x Master /\\ 2.0.0\end{tabular}                     \\
		W-Bridge                                                  & \begin{tabular}[c]{@{}l@{}}Kunbus,\\ IOLW Stack\end{tabular}                            & \begin{tabular}[c]{@{}l@{}}IOLW-Bridge /\\ 2.0.0 (adapted)\end{tabular}             \\
		W-Device                                                       & \begin{tabular}[c]{@{}l@{}}TI,\\ IOLW Stack\end{tabular}                                & \begin{tabular}[c]{@{}l@{}}LAUNCHXL-CC2650 / \\ 2.0.0\end{tabular}                  \\
		smartlight                                                   & Balluff                                                                                 & BNI0085                                                                             \\
		e-stop                                                       & Pilz                                                                                    & PITgatebox IOLS                                                                     \\
		light barrier                                                & Pilz                                                                                    & PSENopt II IOLS                                                                     \\
		Engineering PC                                               & CODESYS IDE                                                                             & 3.5 SP 18                                                                           \\ \hline
	\end{tabular}
	\label{tab:bom}
\end{table}

\subsection{Shop floor} 
The shop floor presents a process that transports goods via two opposing conveyor belts and robots move them from one to the other. An occupancy sensor at the end of each belt and a watchdog timer are monitoring the process. If goods are missing, a feeder fills up the cycle. Safety mechanisms like a light curtain and an emergency stop (e-stop) are implemented to trigger the safe state in case of intended or unintended human intervention. The cell continues with operation by releasing the e-stop and clearing the light barrier. Each sensor and actuator is connected via \ac{iolw} to a W-Master, communicating with the 5G network through a 5G router, which is depicted in Fig.~\ref{fig:testbednetworkarchitecture}. The second part is a decentralized software-based \ac{plc} running on a \ac{revpi} that is also connected via a 5G router. An IPSec \ac{vpn} tunnel bridges the two routers.

\subsection{PLC} 
An industrialized Raspberry Pi, called \acl{revpi}, is used as a \ac{plc}. On the \ac{revpi}, a software based \ac{plc} runtime controls the I/Os on the shop floor. The \ac{plc} is connected to a 5G router (Fig.~\ref{fig:testbednetworkarchitecture}). Through the \ac{vpn} tunnel, the \ac{plc} communicates with the W-Master using the 5G network as if being directly connected. The task cycle time is configured at 5\,ms to assure the data evaluation, whereas the proven reliable \ac{iolw} query cycle is 10\,ms.

\subsection{Network} 
The network is segmented into two parts, one for the controller and the other for the shop floor, as illustrated in the lower part of Fig.~\ref{fig:testbednetworkarchitecture}. The traffic is securely tunneled and routed through the 5G network via an IPSec \ac{vpn} tunnel. The 5G routers are assigned \ac{ip} addresses via \ac{dhcp} on the cellular interface while providing the underlying network on the wired interface. To provide quasi static \ac{ip} addresses in the \ac{dhcp}-managed 5G network, two separate slices are configured to represent a singular \ac{ip} address contingent, depending on the SIM card.

\subsection{5G Configuration} 
The 5G network employs a hybrid \ac{sa} architecture with Release 16 features. The 5G core is distributed between the vendor's site and the university campus, separating the control and management plane from the user plane. This enables administration of the control plane by the vendor on demand. However, all private network traffic remains securely on-premise. The network spans the entire 100\,MHz from 3.7 to 3.8\,GHz, using a \ac{scs} of 30\,kHz. The testbed setup is static, ensuring line-of-sight between all devices and the nearest active antenna unit being approx. 7\,m away. Throughout the testing phase, the \ac{rssi} consistently registered at at least -60 dBm.

\subsection{\acl{iolw} Configuration } 
\begin{figure}[tb]
	\centering
	\resizebox{6.5cm}{!}{
	\begin{tikzpicture}[>=latex']
		\tikzset{block/.style= {draw, rectangle, align=center, minimum width=1cm, minimum height=0.5cm},}
		\node [block] (start) {W-Master};
		
		\node [coordinate, right = 0.35cm of start] (ADL){};
		
		\node [coordinate, above = 2.45cm of ADL] (AUL){};
		\node [coordinate, above = 1.75cm of ADL] (BUL){};
		\node [coordinate, above = 1.05cm of ADL] (CUL){};
		\node [coordinate, above = 0.35cm of ADL] (DUL){};
		\node [coordinate, below = 0.35cm of ADL] (EUL){};
		\node [coordinate, below = 1.05cm of ADL] (FUL){};
		\node [coordinate, below = 1.75cm of ADL] (GUL){};
		\node [coordinate, below = 2.45cm of ADL] (HUL){};
		
		\node [block, right = 0.5cm of AUL] (A1){W-Device};
		\node [block, right = 0.5cm of BUL] (B1){W-Device};
		\node [block, right = 0.5cm of CUL] (C1){W-Device};
		\node [block, right = 0.5cm of DUL] (D1){W-Device};
		\node [block, right = 0.5cm of EUL] (E1){W-Device};
		\node [block, right = 0.5cm of FUL] (F1){W-Bridge};
		\node [block, right = 0.5cm of GUL] (G1){W-Bridge};
		\node [block, right = 0.5cm of HUL] (H1){W-Bridge};

		\node [block, right = 0.5cm of A1] (A2){robot 1};
		\node [block, right = 0.5cm of B1] (B2){robot 2};
		\node [block, right = 0.5cm of C1] (C2){conveyor belt 1};
		\node [block, right = 0.5cm of D1] (D2){conveyor belt 2};
		\node [block, right = 0.5cm of E1] (E2){feeder};
		\node [block, right = 0.5cm of F1] (F2){smart light (IOL)};
		\node [block, right = 0.5cm of G1] (G2){e-stop (IOLS)};
		\node [block, right = 0.5cm of H1] (H2){light barrier (IOLS)};
					
		\path[draw, -]
		(start) -- (ADL)
		(ADL) -- (AUL)
		(ADL) -- (BUL)
		(ADL) -- (CUL)
		(ADL) -- (DUL)
		(ADL) -- (EUL)
		(ADL) -- (FUL)
		(ADL) -- (GUL)
		(ADL) -- (HUL)
		
		(AUL) -- (A1)
		(BUL) -- (B1)
		(CUL) -- (C1)
		(DUL) -- (D1)
		(EUL) -- (E1)
		(FUL) -- (F1)
		(GUL) -- (G1)
		(HUL) -- (H1)
		
		(A1) -- (A2)
		(B1) -- (B2)
		(C1) -- (C2)
		(D1) -- (D2)
		(E1) -- (E2)
		(F1) -- (F2)
		(G1) -- (G2)
		(H1) -- (H2)
	
		;
	\end{tikzpicture}
	}	
	\caption{\ac{iolw} testbed structure.}
	\label{fig:iolw}
\end{figure}
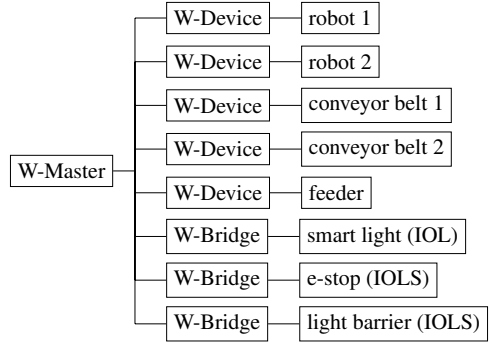

The \ac{iolw} system, shown in Fig.~\ref{fig:iolw}, resp. the W-Master, is configured to operate two tracks with a cycle time of 5\,ms. Track one includes the W-Devices for robots, belts and the feeder, all implemented on \ac{iolw} development kits. Track two communicates with the wired \ac{iol}-Devices via \ac{iol}/\ac{iolw}\nobreakdash-Bridges (W-Bridges). This setup was selected to showcase the flexibility of the testbed and its suitability for industrial-grade devices. In fact, two W-Bridges are adapted to bridge the \ac{iols}-Devices (e-stop and light barrier), integrating functional safety features as well.

\section{Measurements and Result}\label{sec:measurement} 
While assessing latencies, functional safety applications employing wireless communication are the main perspective, such as the necessary \ac{sfrt} of a robot stopping after the  interruption of a light barrier or after activating a wirelessly connected e-stop. The presence of moving objects within the safety application necessitates the establishment of a reliable and deterministic communication channel.

Latency and performance metrics of the IP network part are measured using the Isochronous Round-Trip Tester and iPerf3. A 20-hour test with 71,757 valid measurements yielded the following results for the 5G network: an average round-trip time between routers of 20.4\,ms, a one-way delay of 10.2\,ms from the router to the Core Network and a throughput of 912\,Mbits/s in the downlink and 110\,Mbits/s in the uplink. The Ethernet link test results in an average latency of 1.2\,ms per connection.

\begin{figure*}[tb]
	\centering
	\includegraphics[width=1\linewidth]{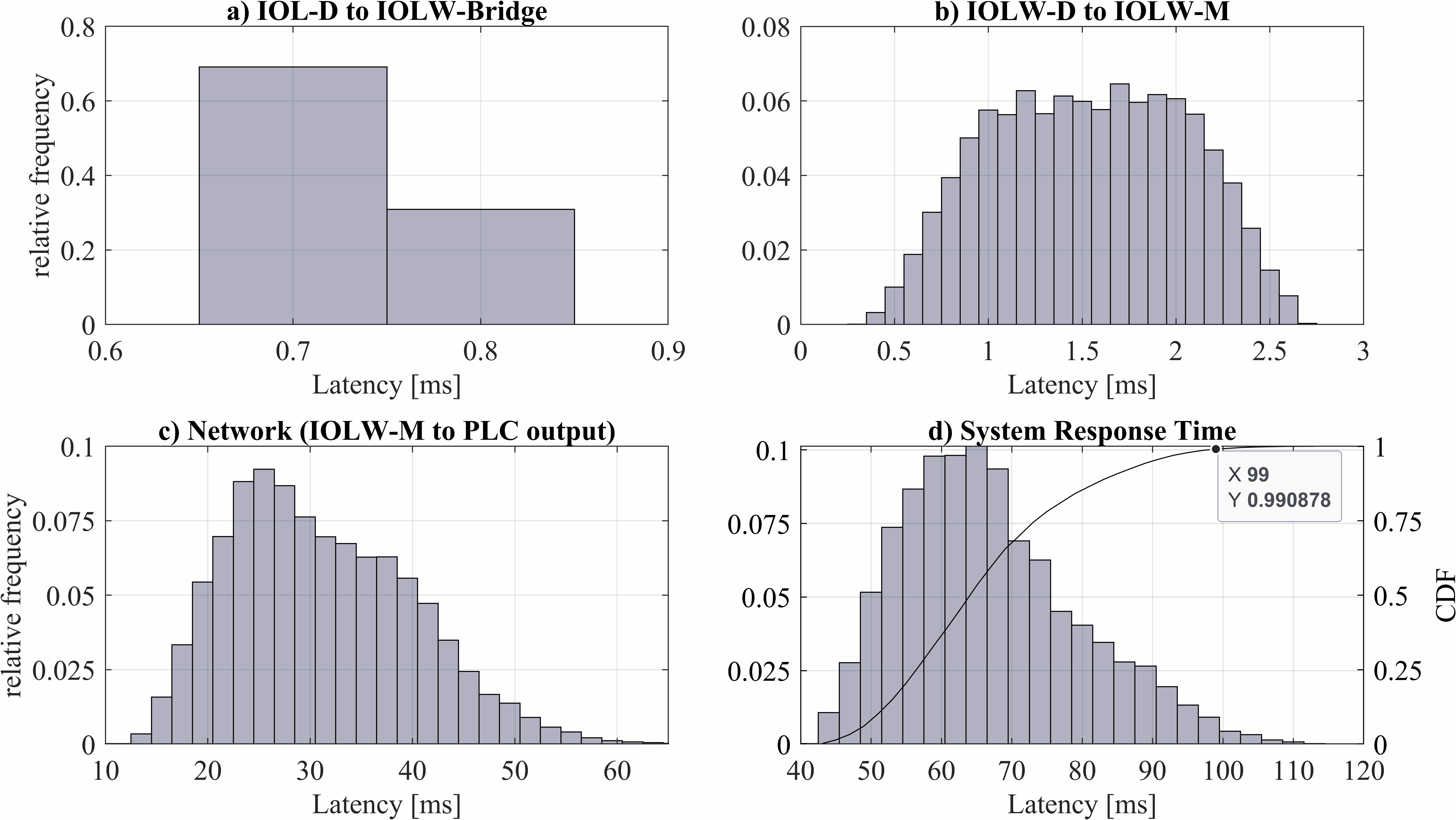}
	\caption{Testbed latencies for a) \ac{iol}, b) \ac{iolw}, c) Ethernet and 5G network and d) the response time of the whole testbed as \ac{sfrt}.}	
	\label{fig:testbedlatencies}
\end{figure*}

Additional measurements were carried out with an oscilloscope connected to the digital output channels of the devices to monitor their communication status. The e-stop, as signal source, is configured to alternate a bit every 200\,ms and the oscilloscope recorded 14,580 valid measurements in 540 sequences, each lasting 5\,s, being sampled at 10\,kS/s. The result is depicted in Fig. \ref{fig:testbedlatencies}. Part a) details the latency on the wired path between \ac{iol}-Device and W-Bridge with a mean latency of 0.7\,ms and part b) the wireless path between W-Device/W-Bridge and W-Master with 1.5\,ms on average. This is quicker than the duration of an \ac{iolw} cycle because changes in process data are transmitted immediately with the subsequent sub-cycle. Fig.~\ref{fig:testbedlatencies} c) shows the probability distribution of transmission times from the W-Master to the digital output of the \ac{plc}, including two Ethernet connections, two 5G links and the \ac{plc} processing. This accumulates to 31.1\,ms on average, representing the system response time for the mentioned path. The delay spread can be attributed to stochastic variations in the timing of the TCP protocol, the different communication cycles and the \ac{plc} process cycle. For instance, if a new value is received just after a \ac{plc} cycle commences, it can only be processed in the following one, causing an additional delay of one cycle.

The average response time for the entire system, e.g., from the e-stop activation to the \ac{plc} processing and back to the robot or smart light, is 66.8\,ms with the distribution of the relative frequency shown in Fig.~\ref{fig:testbedlatencies} d) as well as the cumulative distribution function (CDF) on the right y-axis. A marker indicates that over 99\,\% of function triggers are recognized, transmitted, evaluated and executed by the testbed in under 99\,ms, with a maximum observed latency of 113.5\,ms. Summing the maximum latencies measured for each system segment yields a worst-case estimation of 149.6\,ms. In a safety-related context, the worst-case estimation represents the \ac{sfrt} of the e-stop or light barrier and results in a minimum safety distance of 0.3\,m from the moving parts based on a defined hand motion speed limit of 2\,m/s \cite{noauthor_safety_2010}.

\section{Conclusion and outlook}\label{sec:conclusion} 
The implementation of an automation testbed primarily utilizing a wireless communication infrastructure is introduced as a platform for demonstrative applications. The protocols used can be extensively adapted and exchanged, thereby enabling comparative analyses and optimization for functional and operational safety applications. Future work may focus on the deployment of \ac{iolw} Safety on the shop floor level and ProfiSafe or OPC UA Safety (over 5G) on the field bus level to evolve into a comprehensive safety testbed. Furthermore, latency measurements are presented to verify the suitability of the proposed combination of 5G and \ac{iolw} in an industrial environment with response times in the order of 100\,ms and a \ac{sfrt} of 150\,ms in safety-related applications. Powerful servers as part of the 5G infrastructure at the industrial edge present an opportunity to investigate the impact and performance of a virtual \ac{plc}, potentially decreasing the response times by the latency of one 5G connection.

\section*{Acknowledgement}
We would like to thank our project partner, Telekom Deutschland GmbH, and our technology provider, Ericsson GmbH, for their consistently good support and consistently fast responses.

\section*{Funding}
This research is funded by dtec.bw – Digitalization and Technology Research Center of the Bundeswehr. dtec.bw is funded by the European Union – NextGenerationEU (project “Digital Sensor-2-Cloud Campus Platform” (DS2CCP) with the \href{https://dtecbw.de/home/forschung/hsu/projekt-ds2ccp}{project website}).

\bibliographystyle{unsrt}
\bibliography{literature.bib}

\end{document}